\documentclass[nofootinbib,twocolumn,superscriptaddress]{revtex4-1}
\usepackage{graphicx,epsfig}
\usepackage{graphics,dcolumn,bm,epic, eepic,fleqn,float}
\usepackage{amssymb,amsmath,amsfonts,multirow,rotate,color,xcolor}
\usepackage{soul}
\usepackage{url}

\usepackage{subfigure}

\begin{document}
\author{Chlo\"e Brown}
\affiliation{Computer Laboratory, University of Cambridge, Cambridge (UK)}

\author{Neal Lathia}
\affiliation{Computer Laboratory, University of Cambridge, Cambridge (UK)}

\author{Anastasios Noulas}
\affiliation{Computer Laboratory, University of Cambridge, Cambridge (UK)}

\author{Cecilia Mascolo}
\affiliation{Computer Laboratory, University of Cambridge, Cambridge (UK)}

\author{Vincent Blondel}
\affiliation{Louvain School of Engineering, Universit\'e Catholique de Louvain, Louvain-la-Neuve (Belgium)}
\date{\today}

\begin{abstract}We analyze two large datasets from technological networks with location and social data: user location records from an online location-based social networking service, and anonymized telecommunications data from a European cellphone operator, in order to investigate the differences between individual and group behavior with respect to physical location. We discover agreements between the two datasets: firstly, that individuals are more likely to meet with one friend at a place they have not visited before, but tend to meet at familiar locations when with a larger group. We also find that groups of individuals are more likely to meet at places that their other friends have visited, and that the type of a place strongly affects the propensity for groups to meet there. These differences between group and solo mobility has potential technological applications, for example, in venue recommendation in location-based social networks.\end{abstract}
\title{Group colocation behavior in technological social networks}
\maketitle
\section*{Introduction}
In today's technologically connected world, a large volume of data is available about the social ties between individuals and also about their location. The study of these spatially embedded social networks has been a recent topic of interest for researchers, who have largely studied cellphone datasets~\cite{Onnela07:Structure,Gonzalez08:Understanding,Lambiotte08:Geographical,Krings09:Urban,Onnela11:Geographic} and online social networks~\cite{Backstrom10:Find,Chang11:Location,Cheng11:Exploring,Cho11:Friendship,Cranshaw10:Bridging,Scellato11:Socio}. Datasets of mobile phone calls with cell tower locations, and those from online social networks with real-time indication of their users' whereabouts, available through such means as geo-tagged tweets in Twitter and through user `check-ins' in location-based social networks (LBSNs) such as Foursquare, afford the opportunity to study the relationship between the geographic positions of users and their friends at particular points in time, rather than using static locations such as home addresses. This enables us to investigate the relationship between friendship and colocation: the characteristics of the places where people meet with their friends, and how this may differ from the mobility of a lone individual.

To date, there exists very little research examining the characteristics of the mobility of groups of friends in these networks. Some first work includes that by Crandall \emph{et al.}~\cite{Crandall10:Inferring}, who analyzed how colocation affected the probability that two users of the online photo-sharing service Flickr were friends, showing that even a small number of colocations between users is a strong predictor of a social link. A further study concerning this topic was done by Cranshaw \emph{et al.}~\cite{Cranshaw10:Bridging}, who examined the locations where people were colocated, and developed a measure of the entropy of a place in terms of the variety of unique visitors to that place. They observed that colocations at low entropy places such as homes are more likely to be between friends than those at high entropy places such as a shopping mall or university, which suggests that there may indeed be differences between the places that people visit with their friends and those where they go alone, and thus differences between individual and group mobility patterns. Calabrese \emph{et al.}~\cite{Calabrese11:Interplay} studied directly the interplay between face-to-face interactions and mobile phone communication, finding that colocations appear indicative of coordination calls, which occur just before face-to-face meetings.

This study aims to build on this initial research and study whether there are differences between individual mobility and mobility in social groups, as viewed through the lens of technological networks such as the mobile phone communication network and location-based online social networks. We study two datasets, one cellphone dataset from a European mobile operator, and one from Foursquare, a popular online location-based social networking service, and examine the behavior of colocated friends, compared to the general behavior seen in each dataset. Knowledge of differences between the places where people go by themselves and those they visit when with their friends could be useful to the designers of location-aware social technologies, for example, by providing venue recommendations in location-based social networks, or better location prediction for location-aware advertising or search results returned on mobile phones~\cite{DeDomenico13:Interdependence}. Please note that in this paper we use the word `friend' to indicate `people who communicate with one another using the technological service being analyzed. We do not place restrictions on tie strength to define the social network, although the effect of tie strength on the results remains a potential direction for interesting future study.

\section*{Results}
\subsection*{Behavior of friend pairs in Foursquare}
We first study the places visited by individuals and by pairs of friends as recorded by their check-ins on Foursquare. Foursquare is an online location-based social network (LBSN), where users of the service may connect to their friends, and indicate their location by `checking in' using an application on their mobile phones. We analyze a set of more than 2 million check-ins made by over 100,000 Foursquare users in New York over a period of 10 months (see Methods). Each check-in consists of the ID of the user who made the check-in, the venue to which the check-in was made (a specific place, as opposed to only co-ordinates), and a timestamp. We use in our analysis the public online social network indicated by the users.

\subsubsection*{Social check-ins}
We define a \emph{social check-in} to be one where a user can be assumed to have been colocated at a venue with one of their friends in the social network. Foursquare does not provide `check-out' information, so we consider a pair of friends to have been colocated when they have checked in to the same venue within one hour of one another. 

\begin{figure}
\begin{center}
    \includegraphics[width=\columnwidth]{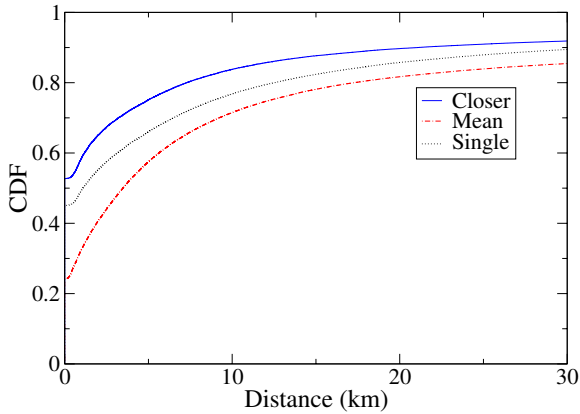}    
\end{center}
  \caption{\textbf{Cumulative distribution function (CDF) of the distance from Foursquare users' top locations to a check-in venue.} Top locations are the locations where a user has checked in the most in the past. `Closer' refers to the distance to the closest of a pair of users' top locations in a colocation event, `mean' refers to the mean distance to the users' top locations, and `single' is the function for all check-ins, not just `social' check-ins that make up colocations between friends. Social check-ins tend to take place closer to a top location of one of a pair of colocated friends than do general check-ins to a top location of the checking-in user.}
  \label{fig:distance_pairs} 
\end{figure}

We analyze the distance of the locations of check-ins from a user's most frequently visited location. We define a user's \emph{top location} to be the Foursquare venue where they have previously checked in the greatest number of times, and compute the distance between this location and the venue where a check-in takes place. We do not consider a user's first check-in in the dataset, so such a location is always defined. In the case that a user has more than one top location, we consider all of those locations, and compute both the mean distance from the check-in venue to those locations, and the minimum distance from the check-in venue to one of those locations. Figure \ref{fig:distance_pairs} shows the cumulative distribution function (CDF) of the distance of a social check-in venue from a user's top locations, and for comparison, the distribution of the distance of the venue of all check-ins (not just social check-ins) to the closest of the top locations of the user making the check-in. The results show that social check-ins tend to be closer to a top location of one of the friend pair concerned than do check-ins in general.

We use information from Foursquare about the categories of venues to analyze the types of places where users tend to go with their friends. Foursquare defines 10 broad categories for venues: Arts and Entertainment (e.g.~theaters, music venues), College and University (e.g.~schools, university buildings), Food (e.g.~cafes, restaurants), Nightlife (e.g.~bars, clubs), Outdoors and Recreation (e.g.~parks, nature spots), Professional (e.g.~workplaces), Residence (e.g.~homes),  Shop and Service (e.g.~shops, hospitals, churches), and Travel and Transport (e.g.~railway stations, airports). We compute the ratio of the proportion of social check-ins in each category to the proportion of all check-ins in that category. Formally, define:
\begin{itemize}
	\item $num\_checkins(c)$ to be the total number of check-ins to category $c$ in the dataset. 
	\item $num\_social\_checkins(c)$ to be the number of social check-ins to category $c$. 
	\item$cats$ to be the set of all categories defined by Foursquare.
\end{itemize}

Then for each category $c$ in $cats$:
\begin{equation*}
	proportion(c) = \frac{num\_checkins(c)}{\sum_{c' \in cats}num\_checkins(c')}
\end{equation*}

That is, the proportion of all check-ins that occur in category $c$, and

\begin{equation*}
	social\_proportion(c) = \frac{num\_social\_checkins(c)}{\sum_{c' \in cats}num\_social\_checkins(c')}
\end{equation*}

That is, the proportion of all social check-ins that occur in category $c$. We then quantify the propensity of each category to include venues where social check-ins particularly take place by defining the colocation ratio:
\begin{equation*}
	colocation\_ratio(c) = \frac{social\_proportion(c)}{proportion(c)}
\end{equation*}

That is, the ratio of the proportion of social check-ins taking place in that category to the proportion of all check-ins taking place in that category. If this value is 1, it means that the category in question is equally likely to host both solo and social check-ins. If the value is markedly less than 1, the category is less likely to host social check-ins than it is to host solo check-ins, so places in this category might be less good to recommend for visits by pairs of friends. If the value is much more than 1, the category is more likely to host social check-ins than it is to host solo check-ins.

\begin{figure}
\begin{center}
    \includegraphics[width=\columnwidth]{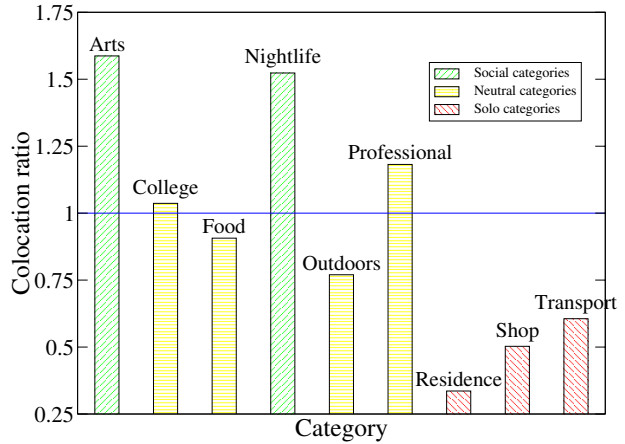}    
\end{center}
  \caption{\textbf{Ratio of the proportion of `social' check-ins (part of a colocation between friends) in each category to the proportion of all check-ins in that category, for each of the categories of venue defined by Foursquare.} Red bars show categories where social check-ins are under-represented (ratio $<0.75$), yellow bars those where social check-ins are approximately in the same proportion as solo check-ins (ratio $0.75 - 1.25$), and green bars show categories where social check-ins are over-represented (ratio $>1.25$). Social check-ins are particularly likely to take place at venues in the Arts and Nightlife categories, and particularly unlikely to take place at venues in the Residence, Shop, and Transport categories.}
    \label{fig:categories} 
  \end{figure}

We compute $colocation\_ratio(c)$ for each category $c$. Figure \ref{fig:categories} shows that more than 1.5 times the proportion of social check-ins are to venues in the Arts and Entertainment and Nightlife Spot categories than the proportion of check-ins in general that are in these categories. Meanwhile, less than 0.7 of the proportion of social check-ins are to Residence, Shop and Service, and Travel and Transport venues than the proportion of check-ins in general.
  \begin{figure}
\begin{center}
    \includegraphics[width=\columnwidth]{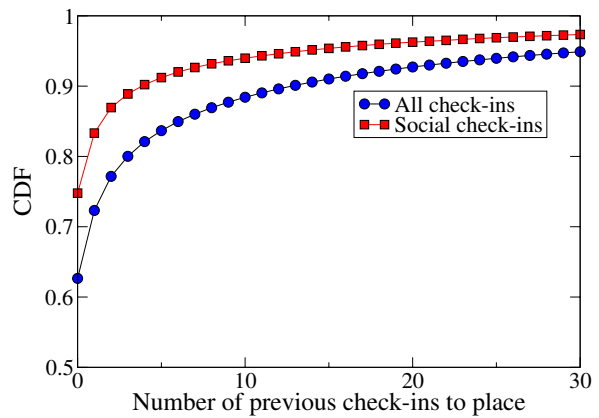} 
\end{center}   
  \caption{\textbf{Cumulative distribution function (CDF) of the number of previous check-ins by a Foursquare user to the check-in venue.} The figure shows the functions for `social' check-ins (part of a colocation between friends) and for all check-ins. Social check-ins are more likely to take place at new venues than check-ins in general.}
    \label{fig:historical_pairs} 
  \end{figure}
\begin{figure}
\begin{center}
    \includegraphics[width=\columnwidth]{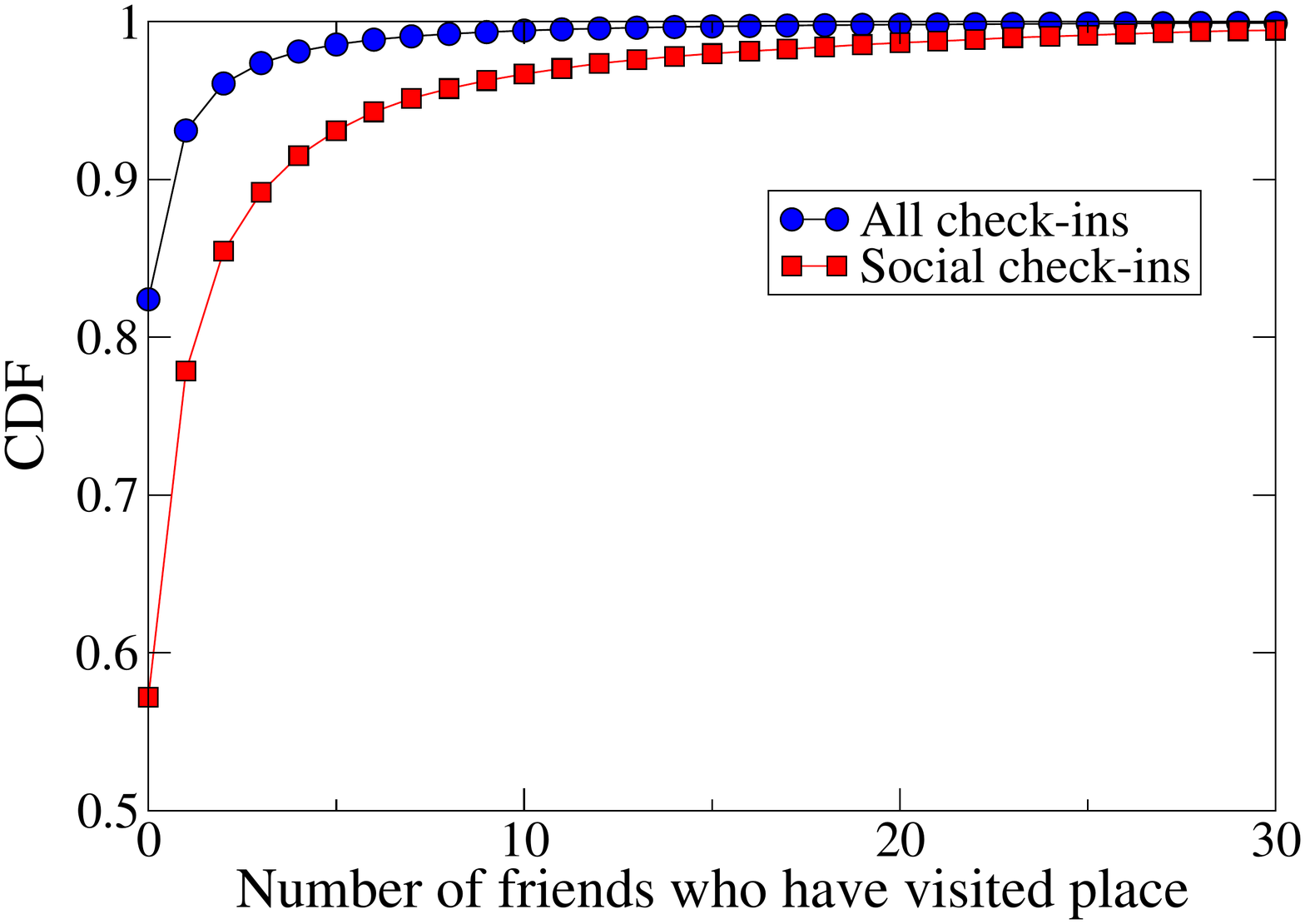}
\end{center}    
  \caption{\textbf{Cumulative distribution function (CDF) of the number of Foursquare users' friends who have previously checked in at the check-in venue in the dataset.} The figure shows the functions for `social' check-ins (part of a colocation between friends) and for all check-ins. Social check-ins are more likely to take place at venues where at least one of the user's friends has been than check-ins in general.}
    \label{fig:friends_pairs} 
  \end{figure}
We examine how likely users are to visit new places with their friends. Figure \ref{fig:historical_pairs} shows the cumulative distribution function (CDF) of the number of previous visits in the dataset by the checking-in user to the visited venue, for all check-ins and for social check-ins. Social check-ins are more likely to take place at new venues than check-ins in general; the probability that a social check-in takes place at a previously visited venue is about 0.25, compared to about 0.38 for any check-in.

Finally, we investigate how many social check-ins take place at venues where members of the friends' wider social circles have previously checked in. Figure \ref{fig:friends_pairs} shows the cumulative distribution function (CDF) of the number of a user's friends who have previously checked in to the venue in question, for all check-ins and for social check-ins (note that we require previous check-ins by friends to be at least an hour before the check-in in question, to avoid counting the first social check-in in a pair constituting a colocation event). Social check-ins are more likely than general check-ins to take place at a venue where a user's friends have been before. About 18\% of all check-ins are to a place visited by at least one friend before, but about 43\% of social check-ins being to such venues. At first glance, this may seem contradictory to the observation that social check-ins tend to take place at venues where the user has not been before, but in fact it is the case that pairs of friends tend to check in together at venues that are new to the pair in question, but not to their wider social circle. It tends to be not one of the pair that has checked in to the venue before, but other friends of one or both of the colocated friends.

\subsection*{Behavior of friend groups in a cellphone network}
We extend our analysis from the behavior of colocated pairs of friends to larger friendship groups. Due to data sparsity, we are not able to obtain meaningful results for larger groups from the Foursquare dataset. Instead, we use a large anonymized dataset of billing records for over one million mobile phone users in Portugal, covering twelve months in 2006 and 2007 (see Methods). We extract colocated friendship groups by first constructing the social network by representing users by nodes and placing edges between them whenever one has called the other. We consider as colocated people who make calls using the same cell tower within one hour of one another, in agreement with the definition of social check-ins in the Foursquare dataset. Within a temporal window of one hour, we consider as groups the connected components of the subgraph of the social network that contains only edges between people colocated during that hour.

\begin{figure}
\begin{center}
    \includegraphics[width=\columnwidth]{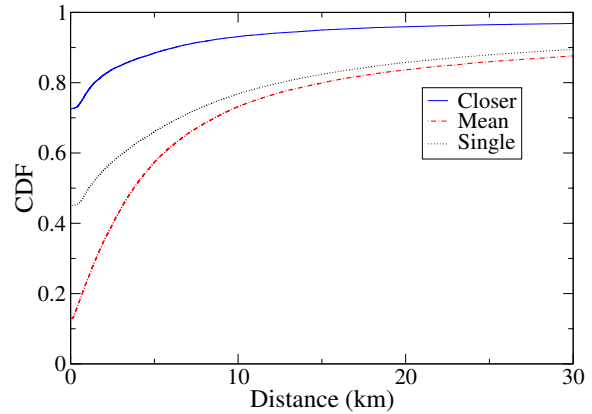}    
\end{center}
  \caption{\textbf{Cumulative distribution function (CDF) of the distance from cellphone users' top locations to a group colocation venue.} Top locations are the locations where a user has made or received a phone call the most in the past. `Closer' refers to the distance to the closest of a user's top locations in a colocation event, `mean' refers to the mean distance to the user's top locations, and `single' is the function for all calls, not just colocations. Groups tend to congregate closer to their members' top locations than people tend to go in general in relation to their top locations.}
  \label{fig:distance_groups} 
\end{figure}

We examine the distance of the places where a person meets a group of their friends from the top location of that person, defined for the cellphone data as the location of the cell tower from which that user has called the most frequently in the past. Figure \ref{fig:distance_groups} shows the cumulative distribution function of the distance of the colocation of a cellphone user with a group of their friends from that userÕs top locations. The figure also shows the distribution of the distance of all calls from the caller's top locations (not just social colocations). Similarly to social check-ins in Foursquare, group colocations are more likely to take place near to one of the group's top locations than calls in general. We note that the higher values of the CDF at very small distances compared to Foursquare is due to the coarser-grained spatial resolution of the cellphone dataset; in Foursquare we have individual venues for check-ins, giving single-building accuracy, whereas in the cellphone dataset we are limited to the resolution of cell towers.

\begin{figure}
\begin{center}
    \includegraphics[width=\columnwidth]{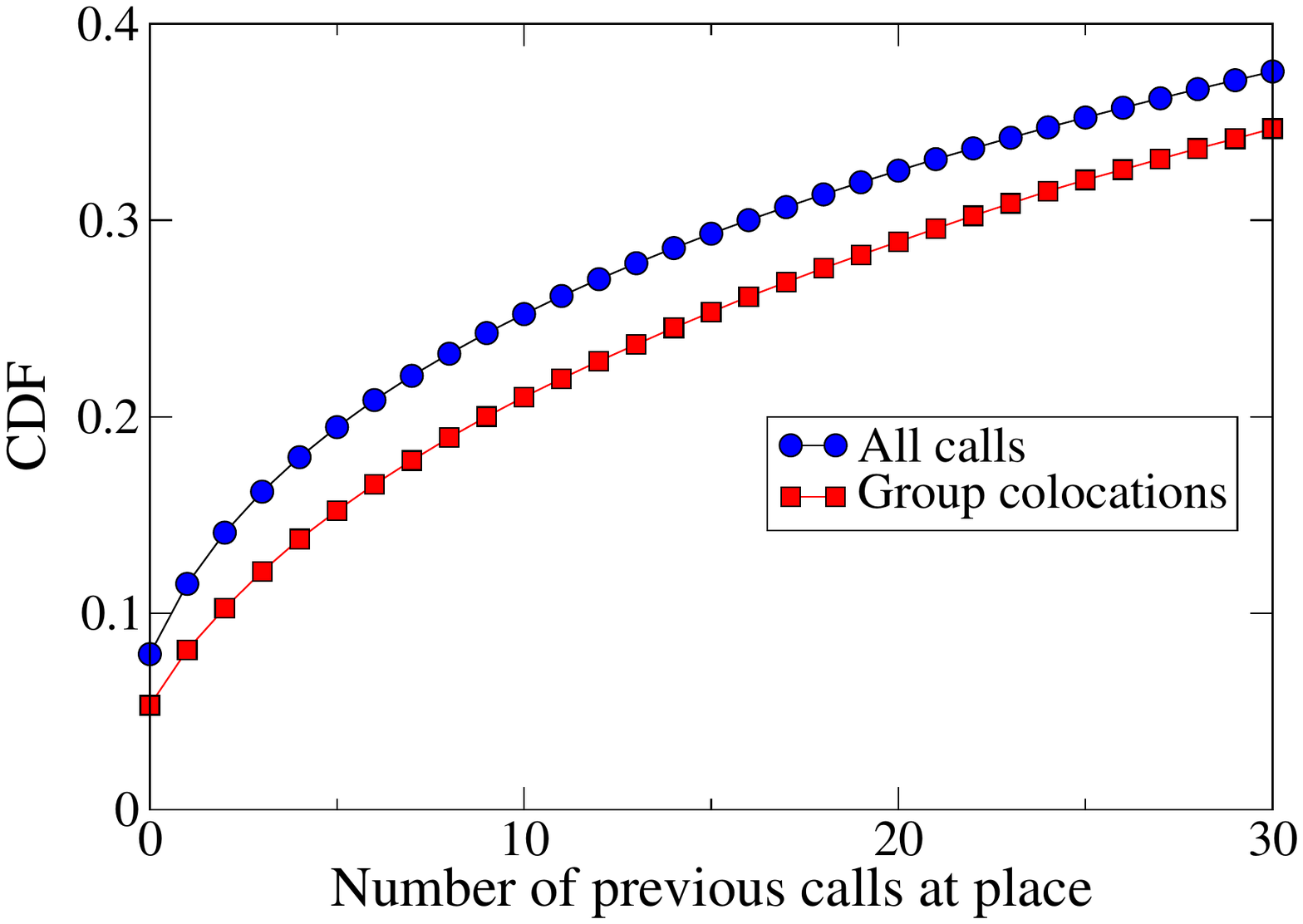} 
\end{center}   
  \caption{\textbf{Cumulative distribution function (CDF) of the number of previous calls made or received by a cellphone user at a location.} The figure shows the functions for group colocations and for all calls. Group colocations are less likely to take place at new places, contrary to what was seen for pairs in the Foursquare data.}
    \label{fig:historical_groups} 
  \end{figure}

\begin{figure}
\begin{center}
    \includegraphics[width=\columnwidth]{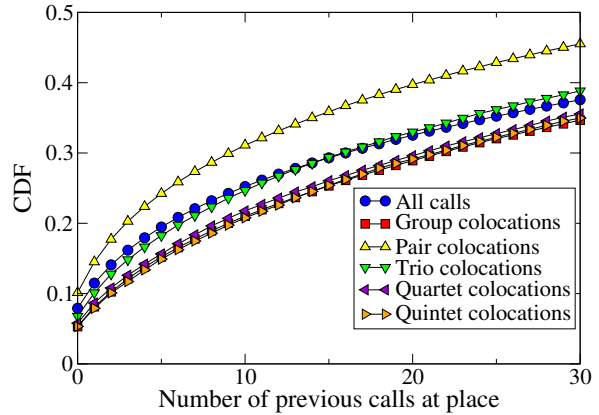} 
\end{center}   
  \caption{\textbf{Cumulative distribution function (CDF) of the number of previous calls made or received by a cellphone user at a location, for colocated groups of various sizes.} Pairs are more likely to meet at new places, in agreement with the Foursquare data, but bigger groups than this tend to meet at familiar locations.}
    \label{fig:historical_breakdown} 
  \end{figure}

Figure \ref{fig:historical_groups} shows the CDF of the number of times a member of a colocated group has previously been seen at the place of colocation. Contrary to the behavior of pairs in Foursquare, an individual is less likely to meet a group of their friends at a new place than at a place where they have been before. We investigate this phenomenon further by analyzing separately groups of different sizes: pairs, trios, quartets, and quintets. Figure \ref{fig:historical_breakdown} shows the CDF for groups of each size. The results are in agreement with the results from Foursquare: pairs are more likely to meet at new places than people are to call from new places in general, but an individual is likely to meet a larger group somewhere they have been before.

\begin{figure}
\begin{center}
    \includegraphics[width=\columnwidth]{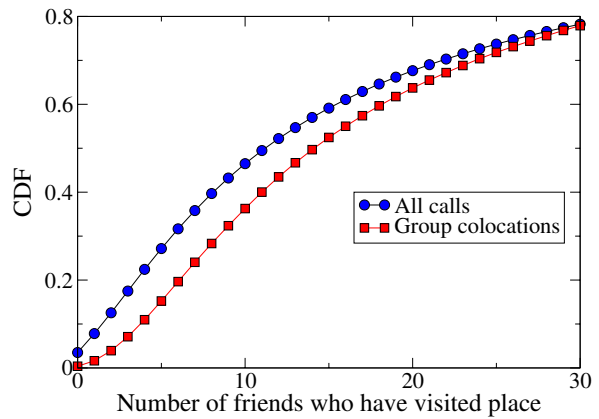}
\end{center}    
  \caption{\textbf{Cumulative distribution function (CDF) of the number of cellphone users' friends who have previously made a call from a location.} Group colocations are more likely to take place somewhere where at least one of the user's friends has been than calls in general.}
    \label{fig:social_groups} 
  \end{figure}

Finally, Figure \ref{fig:social_groups} shows the CDF of the number of an individual's friends, excluding the group with whom they are colocated, who have been at the colocation place before. Again in agreement with the Foursquare data, a user is likely to meet a group at places where their wider circle of friends have been previously, compared to somewhere where none of their friends have been.
  
\section*{Discussion}
In this study we examine the differences between individual mobility and group colocation in two large datasets, one from the online location-based social network Foursquare, and one from a year's worth of CDRs from a European cellphone operator. We find that there are indeed differences; first of all that in both datasets, friends are likely to meet closer to one of the individuals' familiar locations than people may go in general. This is in agreement with the finding by Calabrese \emph{et al.}~\cite{Calabrese11:Interplay} that as the distance between the homes of pairs of colocated users increases, colocations take place in an area closer to one of the homes of the pair. We also analyze the venue category information available from Foursquare, and show that meetings between friends are over-represented in some categories such as Entertainment and Nightlife, and under-represented in others, such as Shop and Transport. This gives some indication that people are more likely to visit some categories of places with their friends than others, which could have implications for place recommendation in online location-based social networks. In both datasets, it is also the case that people are more likely to be colocated with friends at places where their other friends not part of that group have been before.

In analyzing the number of times people have been recorded at a colocation venue previously, we found differing behavior between pairs of friends (both in Foursquare and cellphone dataset) and larger groups: while an individual is more likely to travel to a new place with a single friend than they are on their own, the opposite behavior is seen where larger groups are concerned. One way to interpret this could be in terms of research that has shown that people are more likely to take risks when with peers~\cite{Gardner05:Peer}, possibly including visiting a new place. Being with a friend may increase confidence and willingness to explore. However, this must be balanced, in a larger group situation, with the fact that the larger a group, the more difficult it is for that group to coordinate~\cite{Ingham74:Ringelmann}. This could mean that it is easier for the group to meet at a place familiar to all of its members, than to identify and agree on a new meeting place. 

As with any study of this kind, there are some limitations that must be considered when interpreting the findings; these results should not be used to draw general conclusions about social and mobility behavior. The analysis of the behavior of Foursquare users in this paper is only certain to be relevant to those Foursquare users in the data, who are by their nature as users of the service not representative of a general population, and similarly for the mobile phone users whose behavior is reflected by the telecoms data. In particular, the Foursquare dataset contains check-ins that users shared publicly during the data collection period, and those users may have been to places where they did not check in, or kept certain check-ins private. It must be borne in mind that the likelihood of sharing a check-in at a given place may not be independent of the place category; for example, users might be less likely to share check-ins at home or at a train station than check-ins at entertainment venues. It is also possible that some of the patterns seen in Foursquare check-ins do extend outside the specifics of the services, since building on Cho, Myers, and Leskovec's observation of considerable similarity between the mobility patterns in some LBSN datasets and in mobile telephone network data~\cite{Cho11:Friendship}, we have shown similarities also between the mobility behavior of groups seen in the Foursquare dataset analyzed in this work and the dataset from the mobile network operator. 

Of course, mobile phone datasets also have their inherent biases, and in this case it could be that we underestimate colocation between users because they did not happen to call one another around the time of meeting (similarly to people not checking in on Foursquare) or because they were connected to different cell towers with overlapping coverage areas, or that false positive colocations are inferred because people called one another while connected to the same cell tower despite not meeting face-to-face. We observe that the similarities in behavior that we find between the Foursquare and telecoms datasets provides some evidence in support of the differences not being due to these effects. However, the most reliable applications of results from the analysis of data such as these are most probably in areas directly relating to the use of those services, whether the mobile phone network or online social networks. Taken in the context of the technological service concerned, that is, for applications \emph{within} such services, such as venue recommendation for groups in LBSNs, the results are relevant and potentially useful.

We argue that our findings have potential applications in location-aware technological services. For example, one important task in online location-based social networks is that of venue recommendation: suggesting to the logged-in user places where that user might like to go. Given that we have found differences between the places that people may visit with their friends and those where they may go on their own, it may be that services such as Foursquare could benefit from suggesting different recommendations to a user based on knowledge of whether they are with friends or alone. Similarly, these differences in individual and group behavior could be used to improve mobility prediction, with possible uses in personalized services such as location-aware advertising and search results returned on mobile phones~\cite{DeDomenico13:Interdependence}.

\section*{Methods}
The Foursquare dataset consists of check-ins made by Foursquare users between November 2010 and September 2011. The dataset is composed of information publicly available on the Internet that can therefore be downloaded and analyzed in such a way as we have here. We downloaded all of the check-ins that users posted as non-private on Twitter, and which are thus publicly available, through the official Twitter search API by searching for tweets containing `4sq' and a link (to the check-in page). This resulted in a dataset estimated to account for 20-25\% of Foursquare check-ins made during this period. We were also able to download venue information including the venue location (latitude and longitude) and category (College, Food, Professional, etc.) directly from the Foursquare venue database, which is also publicly accessible.

We analyze the set of 2,315,350 check-ins made to 109,314 venues in New York, by 104,266 users. We use check-ins as the representation of user visits to places; check-ins are voluntary so they are an under-estimate of the actual set of places visited by a user. However, the comparison between check-ins made with friends and alone shows considerable differences which may be reflected in the general place visit behavior of users, since the mobility patterns apparent from LBSN check-ins show strong similarity to those seen in other kinds of mobility data including cellphone records~\cite{Cho11:Friendship}. In addition we downloaded the publicly available social networks of the users, taking a pair to be friends when each follows the other on Twitter. This social network has 99,725 nodes and 821,948 edges.

The telecoms dataset is a large anonymized set of billing records for over one million mobile phone users in Portugal, gathered over a twelve month period between 2006 and 2007. The dataset contains information about mobile phone calls, but does not contain text messages (SMS) or data usage (Internet). The dataset was obtained directly from the operator and full permission was granted for its use in our analysis. In order to preserve privacy, individual phone numbers were anonymized by the operator. Each user in the anonymized dataset is identified by a hashed ID. Each entry is a CDR (Call Detail Record), consisting of a timestamp, the IDs of the caller and the callee, the call duration, and the cell tower IDs of the caller and callee towers. The dataset also includes the latitude and longitude of the cell towers, which allows us to study the relationship between the social network constructed by placing edges between nodes representing people whenever one person has called another, and the physical location of the users. The social network has 1,954,188 nodes and 19,370,004 edges.

\section*{Acknowledgements}
Chlo\"e Brown is a recipient of the Google Europe Fellowship in Mobile Computing, and this research is supported in part by this Google Fellowship. We acknowledge the support of the Engineering and Physical Sciences Research Council through grants GALE (EP/K019392) and UBHAVE (EP/I032673/1).

\end{document}